\PassOptionsToPackage{square, comma, sort&compress}{natbib}
\documentclass[preprint,12pt]{elsarticle}

\usepackage{amsmath}
\usepackage[colorlinks=true,linkcolor=blue,urlcolor=blue,citecolor=blue]{hyperref}
\usepackage{graphicx}
\usepackage[squaren, Gray, cdot]{SIunits}
\usepackage{amssymb}

\usepackage{tikz}

\journal{Physics Letters A}

\begin{document}

\begin{frontmatter}



\title{Features of alkali D$_2$ line magnetically-induced transitions excited under $\pi$-polarized laser radiation
}


\author[IPR]{Armen Sargsyan}
\author[FEMTO]{Emmanuel Klinger\corref{mycorrespondingauthor}}
\cortext[mycorrespondingauthor]{Corresponding author}
\ead{emmanuel.klinger@femto-st.fr}
\author[IPR]{Arevik Amiryan}
\author[IPR]{David Sarkisyan}

\address[IPR]{Institute for Physical Research – National Academy of Sciences of Armenia, Ashtarak-2 0203, Armenia}

\address[FEMTO]{Universit\'e Marie et Louis Pasteur, SUPMICROTECH, CNRS, institut FEMTO-ST, F-25000 Besan\c con, France}


\begin{abstract}
The impact of the optical field polarization on the spectrum of magnetically-induced transitions, a class of transitions forbidden at zero magnetic field, is studied with a weak-probe sub-Doppler technique. The high spectral resolution of the technique combined with the simplicity in interpreting the observed spectra allows to follow the behavior of individual transitions as a function of the magnetic field amplitude. We observe only one intense transition (out of $2F_g+1$, where $F$ is the quantum number associated with the total angular momentum of the atom) in the case of $\pi$-polarized laser radiation (a configuration where the applied magnetic field is parallel to the electric field from the laser radiation) in the $F_g\rightarrow F_g+2$ manifolds of $^{85}$Rb, $^{87}$Rb and $^{133}$Cs for fields above a few hundreds of gauss. 
We show that this behavior is in agreement with a model based on the diagonalization of the Zeeman Hamiltonian matrix. With the rapid development of micro-machined vapor-cell-based sensors, these results will be of use to magnetometers operating above Earth field, wide-range laser frequency stabilization systems and atomic Faraday filters.
\end{abstract}




\end{frontmatter}

\section{Introduction}
 
Optical and magneto-optical processes formed in hot atomic vapors of alkali metal atoms are at the core of many devices and applications, such as atomic optical clocks  and gyroscopes \cite{kitching2018chip}, optical magnetometers \cite{FabricantNJP2023}, frequency markers of atomic transitions \cite{pizzey2022laser}, realization of narrow optical resonances based on coherent processes \cite{finkelstein2023practical}, sensitive sensors of electric and microwave fields based on Rydberg atoms \cite{DownesNJP2023}, etc. Therefore, the identification of new intense atomic transitions of alkali metals is of great interest.

The Zeeman effect occurs when atoms are in the presence of an external magnetic field, leading to a significant modification of the intensity of atomic transition \cite{tremblayPRA1990, alexandrov1993interference}. One interesting consequence of the couplings between magnetic sub-levels induced by the Zeeman effect is the possibility to observe $n^2\text{S}_{1/2}\rightarrow n^2\text{P}_{3/2}$ transitions (where $n=3,4,5,6$ is the principal quantum number for Na, K, Rb, Cs, respectively) with an apparent  $\Delta F = \pm 2$ selection rule at non-zero field.
These transitions, referred to as magnetically-induced (MI) transitions, form a large class of one hundred transitions with interesting and important features \cite{sargsyanJPB2020,sargsyanPhysLetA2021,Sargsyan2024observation}. 
Interest in these transitions is primarily driven by the fact that, in wide ranges of magnetic fields, their intensities can significantly exceed the intensities of “usual” (i.e. $\Delta F = 0,\pm1$), widely used atomic transitions. In addition, their frequency shifts in strong magnetic fields reach 20 to 30 GHz, enabling the exploration of new frequency ranges
for the frequency stabilization of lasers operating at significantly shifted frequencies compared to unperturbed atomic transitions. These transitions can also be used to form coherent processes such as electromagnetic-induced transparency 
in large magnetic fields \cite{sargsyan2019dark,sargsyan2022coherent}. A thorough understanding of the evolution of alkali transitions in a magnetic field is also important for magnetometry above Earth field \cite{klingerAO2020,staerkind2023high}.  Note that MI transitions can also appear in spectra of ions \cite{Grumer2014unexpected}, and atoms other than alkali \cite{Winchester2017magnetically}. Similar atomic transitions are also used in solar physics \cite{judge2014forbidden,schad2021polarized}.


Despite possessing interesting features, magnetically-induced transitions are often disregarded because traditional weak-probe spectroscopy lacks the resolution to distinguish individual transitions due to the Doppler effect. Conversely, sub-Doppler spectra acquired through nonlinear techniques, such as saturated absorption, are cumbersome to interpret because of the large number of transitions \cite{scottoPRA2015}. However, with nanocells (NCs), where the vapor is confined between two almost-parallel windows separated by a distance ranging from a few nm to several $\mu$m, one can realize weak-probe sub-Doppler spectroscopy \cite{dutier2003collapse}. By selecting the proper cell thickness, this can be used to realize sub-Doppler spectroscopy of molecular lines \cite{arellanoNC2024}. 
 Beyond spectroscopy, NCs serve as a crucial tool in fundamental physics. They are, for instance, used to study atom-surface interactions \cite{Laliotis2021atom-surface,sargsyanPLA2023,duttaPRS2024}, cooperative effects \cite{peyrot2019Optical}, and realize single-photon generation at room temperature \cite{ripka2018room}. New advancements in manufacturing techniques \cite{PeyrotOL2019} are making nanocells (NCs) increasingly accessible.

In this article, we study the evolution of MI transitions of Rb and Cs vapors, confined in a NC, in the presence of a magnetic field. To improve the resolution of the measurement, second derivative spectroscopy \cite{demtroder1981laser,sargsyanOL2019} is used. We show that the spectrum of $F_g \rightarrow F_g + 2$ MI transitions excited under linearly-polarized ($\pi$) optical field contains only one intense transition out of $2F_g + 1$. In contrast, the spectrum of $F_g \rightarrow F_g + 2$ MI transitions excited under circularly-polarized ($\sigma^+$) optical field contains $2F_g + 1$ intense transitions, that is 3 transitions for $^{87}$Rb, 5 transitions for $^{85}$Rb, and 7 transitions for $^{133}$Cs. 



\section{Alkali atoms in a magnetic field}

In the presence of a magnetic field, magnetic sublevels are split and mixed, leading to changes in transition probabilities and frequencies. These changes can be calculated by diagonalizing the Hamiltonian matrix \cite{tremblayPRA1990}, which accounts for the hyperfine atomic structure, 
and the Zeeman Hamiltonian, that is
\begin{equation}\label{eq:ZeemanMatrix}
    H = H_{\text{hfs}} +\frac{\mu_B}{\hbar}(g_S\,S_z +g_L\,L_z + g_I\,I_z) B_z\,,
\end{equation}
where $\mu_B$ is the Bohr magneton \cite{olsen2011optical}, $g_{S,L,I}$ are the Landé factors \cite{staerkind2023precision} and $S_z$, $L_z$, $I_z$ the projection of quantum numbers on the $z$-axis (quantization axis). The eigenvalues obtained after the diagonalization of the matrix provide access to the resonance frequencies. The transition intensities are proportional to the squared dipole moment, given, in the coupled basis $|F,m_F\rangle$, by 
\begin{equation}
    |\langle e||d||g\rangle|^2 \propto \Gamma_\text{N} \, a^2[\psi(F_e',m_{F_e});\psi(F_g',m_{F_g});q]\,,
\end{equation}
where $\Gamma_\text{N}$ the natural linewidth of the transition, $a[\psi(F_e',m_{F_e});\psi(F_g',m_{F_g});q]$ are the transfer coefficients modified by the field \cite{tremblayPRA1990}, calculated using the eigenvectors. Here, $q=0,\pm1$ is associated with the polarization of the incident laser field. To excite $\pi$ ($q=0$) transitions, the polarization of the electric field of the laser should be collinear with the applied magnetic field axis.

\begin{figure}[htb]
    \centering
    \includegraphics[width=0.55\textwidth]{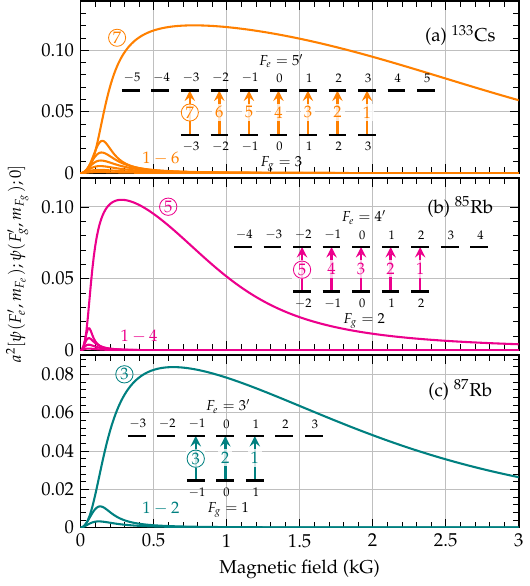}
    \caption{Dependence of the intensity of (a) $^{133}$Cs $6^2\text{S}_{1/2}\rightarrow 6^2\text{P}_{3/2}$, (b) $^{85}$Rb $5^2\text{S}_{1/2}\rightarrow 5^2\text{P}_{3/2}$ and (c) $^{87}$Rb $5^2\text{S}_{1/2}\rightarrow 5^2\text{P}_{3/2}$ $F_g \rightarrow F_g +2$ MI transitions versus magnetic field when excited by linear $\pi$-polarization.  In large magnetic fields, atoms are left with only one MI transition (out of $2F_g+1$). In each plot, the inset shows the diagram of transitions with the corresponding labels.}
    \label{fig:Fig1}
\end{figure}

Figure~\ref{fig:Fig1} shows the calculated evolution of $F_g\rightarrow F_g + 2$ MI transitions of the $D_2$ line of $^{133}$Cs (a), $^{85}$Rb (b) and $^{87}$Rb (c) under $\pi$ optical excitation. A rapid increase of the MI transition intensities can be seen as a function of the magnetic field for all three atoms. The characteristic magnetic field\footnote{$B_0$ is 0.7\,kG, 2.4\,kG and 1.7\,kG for $^{85}$Rb, $^{87}$Rb and $^{133}$Cs, respectively.} $B_0=A_{\text{hfs}}/\mu_B$, where $A_{\text{hfs}}$ is the magnetic dipole constant of the ground state, is a good indicator for the dynamics of transitions. For each atom, one transition is seen to have a much larger increase in intensity, reaching a maximum at about $B \sim 0.5 B_0$. With a further increase of the magnetic field, the intensity smoothly decreases to almost zero. Note that a field $B\gg B_0$ marks the onset of the hyperfine Paschen-Back regime \cite{weller2012absolute}. These transitions are labeled \textcircled{3}, \textcircled{5}, \textcircled{7}, for $^{87}$Rb, $^{85}$Rb and $^{133}$Cs, respectively, see the insets in Fig.\,\ref{fig:Fig1}. Conversely, the intensities of all other transitions quickly tend to zero for fields $B > 0.2 B_0$. 
Thus, for a magnetic field above $0.2 B_0$, only one $\Delta F = +2$ transition remains in the spectrum of alkali D$_2$ line observed under $\pi$-polarization. 

 \begin{figure}[htb]
    \centering
    \includegraphics[width=0.55\textwidth]{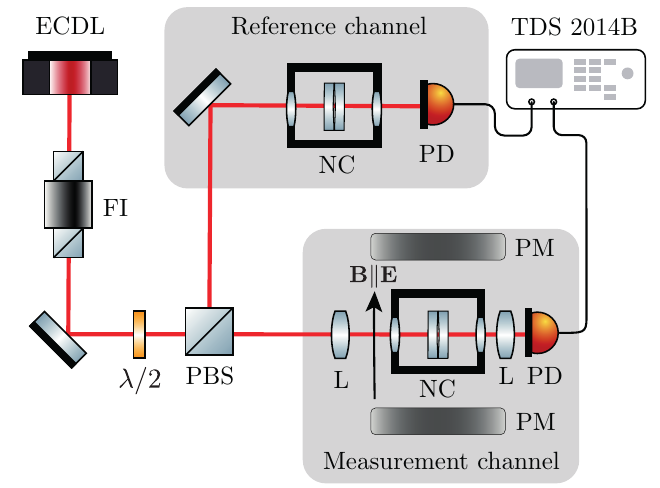}
    \caption{Sketch of the experimental setup. ECDL -- extended cavity diode laser, FI -- Faraday isolator, PBS -- polarizing beam splitter, L -- lens, PM -- permanent magnet, NC filled with Rb or Cs in the oven, PD -- photodiodes. The PMs are oriented such that applied magnetic field \textbf{B} is collinear with the electric field of the laser radiation \textbf{E} (\textbf{B}$\parallel$\textbf{E}) in order to excite $\pi$ transitions. The reference channel comprises a Rb (or Cs) nanocell from which the transmission spectrum of the D$_2$ line is recorded at zero magnetic field.
}
    \label{fig:Fig2}
\end{figure}

\section{Experimental arrangement}
 The sketch of the experimental setup is shown in Fig.~\ref{fig:Fig2}. A linearly-polarized laser beam, emitted by a cw extended cavity tunable diode laser (ECDL) \cite{vassilievRevSciInstrum2006} is used. When the experiments are carried out with Rb atomic vapor, an ECDL with $\lambda =780$ nm is tuned in the vicinity of the Rb $D_2$ line. In the case of Cs atomic vapor, another ECDL with $\lambda=852$ nm, is tuned in the vicinity of the Cs $D_2$ line.  In both cases, the spectral width of the probe laser radiation is $\gamma_l <1$\;MHz. In order to avoid feedback in the cavity, a Faraday insulator is placed immediately after the laser. A polarizing beam splitter (PBS) is used to purify the initial linear optical polarization and direct part of the light to an auxiliary channel serving as a frequency reference. The  laser fields transmitted through NCs in both the measurements and reference channels are detected by a photodiode and the signal fed to a Tektronix TDS 2014B digital four-channel oscilloscope. 

  To excite $\pi$ transitions, the applied magnetic field should be collinear with the polarization of the incident laser field, i.e. $\bold{B} \parallel \bold{E}$. 
  To apply fields above 0.5\,kG, a set of two strong permanent magnets mounted on a translation stage was used. The magnetic field strength applied to the atomic vapor is varied via a simple longitudinal displacement of the PM set \cite{sargsyan2012hyperfine}, calibrated with a Teslameter HT201 magnetometer.  Because, in this configuration, the field is applied along the long side of the nanocell with PMs, a high gradient is also introduced and not mitigated as in the case of $\sigma^\pm$ excitations \cite{klingerAO2020}. This results in an additional broadening of the transitions. To form the narrowest resonance possible, the spectroscopy is performed at a vapor column thickness
 of $\ell = \lambda/2$ \cite{dutier2003collapse}. The laser beam diameter is reduced down to $\sim1\;$mm to only illuminate a region where $\ell$ is homogeneous. The acquired spectra are then post-processed with second derivative (SD). The resulting spectrum $T''(\nu)$ typically shows 
 transitions with spectral linewidth of $\sim 20-30$ MHz, which  is narrower by a factor of $\sim$20 compared with the Doppler broadening.
 This allows one to separate and study atomic transitions individually.
 Note that the SD method correctly shows the relative intensities of atomic transitions and their frequency positions; it can therefore be used for quantitative measurements \cite{sargsyanOL2019}.  


\section{Results}
\begin{figure}[htb]
    \centering
    \includegraphics[width=0.55\textwidth]{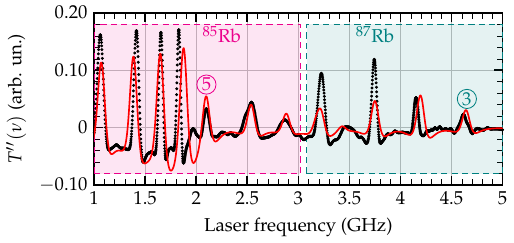}
    \caption{Second derivative transmission spectrum recorded in the vicinity of the $^{85}$Rb $F = 2 \rightarrow 3',4'$ and $^{87}$Rb $F= 1 \rightarrow 2',3'$ manifolds. The experimental spectrum is obtained with a \( \pi \)-polarized excitation at $ B = 800$\;G and $\ell = 390 $\,nm. The red solid line represents the calculated spectrum. The weighted center of the $^{85}$Rb D$_2$ line was chosen as the zero laser frequency. In this spectrum, only one MI transition remains at this field: transition \textcircled{5} (out of a possible five, for $^{85}$Rb), transition \textcircled{3} (out of a possible three for, $^{87}$Rb). }
    \label{fig:Fig3}
\end{figure}

Figure \ref{fig:Fig3} shows the experimental (black dots) SD transmission spectrum obtained using a NC filled with natural-abundance Rb atomic vapor at a column thickness $\ell = 390$\;nm and a field $B=800\;$G, upon excitation by a $\pi$-polarized radiation. It was recorded with a laser power of $20\;\mu$W, and a NC temperature of $120^\circ$C. The laser frequency is scanned across the $^{85}$Rb $F = 2 \rightarrow 3', 4'$ and $^{87}$Rb $F= 1 \rightarrow 2'$, $3'$ manifolds of the D$_2$ line. The red solid line corresponds to the theoretical SD spectrum, calculated by second order numerical differentiation of the transmission spectrum. The theoretical transmission spectrum is obtained by summing the contribution of all possible transitions $j$, each of which is treated as an independent two-level system:
\begin{equation}\label{eq:spectrumT}
    T(\nu) \propto \frac{1}{|Q|^2}\cdot\sum_j\text{Im}\left[\chi_j(\nu,\ell,B)\right] ,
\end{equation}
where $Q=1-r^2\exp(2ik\ell)$, $r$ is the (field) reflection coefficient of the cell windows. This approximation, valid at low laser intensities, is used to speed up calculations. The two-level system lineshape $\chi_j(\nu,\ell,B)$, derived in Ref.\,\cite{dutierJOSAB2003}, is a function of the cell thickness $\ell$ and the laser frequency $\nu$; it reads
\begin{equation}\label{eq:profile}
\begin{split}
    \chi_j(\nu,\ell,B)=-4(1-r\exp[ik\ell])^2&\cdot\frac{\sin^2(k\ell/2)}{Q}\\
    &\cdot\frac{\mathcal{N}}{ku\sqrt{\pi}}\cdot\frac{iA_j(B)}{\Gamma/4\pi-i\Delta_j(B)},
\end{split}
\end{equation}
where $\mathcal{N}$ is the vapor density, $u(\Theta)=\sqrt{2k_B\Theta/m_a}$ is the thermal velocity at a temperature $\Theta$ for atoms of mass $m_a$. The transition parameters $\Delta_j(B)=\nu-\nu_j(B)$  and $A_j(B)$ are fed from the results obtained after diagonalization of the Hamiltonian matrix, see \eqref{eq:ZeemanMatrix}. The homogeneous broadening $\Gamma$, including contributions from natural linewidth, collisional broadening, etc., is left as a free parameter in our simulations. As seen in Fig.\;\ref{fig:Fig3}, only one MI transition can be observed for each atom: \textcircled{5} for $^{85}$Rb and \textcircled{3} for $^{87}$Rb. In similar conditions, $2F_g+1$ transitions have been observed in the case of $\sigma^+$-polarized excitation \cite{klinger2017magnetic}. Such a strong difference in the number of MI transitions excited when using $\sigma^+$ and $\pi$-polarizations is a manifestation of an anomalously strong polarization dependence of the intensity of transitions. Other nearby transitions belong to the $^{85}$Rb $F = 2 \rightarrow 3'$ and $^{87}$Rb $F= 1 \rightarrow 2'$ manifolds. A good agreement between the experiment and the theory can be observed. Nevertheless, some discrepancies can be seen which are likely due to laser instability during the measurements.

\begin{figure}[htb]
    \centering
    \includegraphics[width=0.55\textwidth]{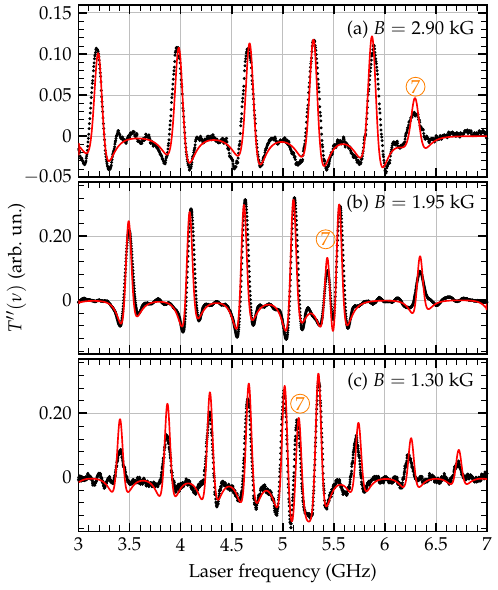}
    \caption{Second derivative transmission spectrum recorded in the vicinity of the $F = 3 \rightarrow 4',5'$ manifolds of the $^{133}$Cs D$_2$ line. The experimental spectra are obtained with a \( \pi \)-polarized excitation, with $\ell = 426 $\,nm and $B = 2.90$\;kG (a), $1.95$\;kG\;(b), $1.30$\;kG (c). The red solid lines represent the respective calculated spectra. The weighted center of the $^{133}$Cs D$_2$ line was chosen as the zero laser frequency. In this spectrum, only one MI transition (out of a possible seven) remains at these fields: transition \textcircled{7}. The theory describes very well the frequency positions and amplitudes of the peaks.     
}
\label{fig:Fig4}
\end{figure}


Figure \ref{fig:Fig4} shows SD transmission spectra upon excitation by a linear $ \pi $-polarized radiation, obtained from a Cs-filled NC at $\ell = 426$\;nm and $B=2.90\;$kG (a), $1.95\;$kG (b) and $1.30\;$kG (c). For the experimental spectra (black dots), the laser is scanned in the vicinity of the D$_2$ line $F = 3 \rightarrow 3', 4', 5'$ transitions with a power of 30 $\mu $W. The NC temperature is set to 100$^\circ $C. 
As the theory predicts, there remains only one MI transition at these magnetic fields, the transition labeled \textcircled{7}, out of the possible 7. The other visible transitions belong to the $F= 3 \rightarrow 3',4'$ of the Cs D$_2$ line manifold. Note that seven MI transitions have been observed in the case of $\sigma^+$-polarized excitation in similar conditions \cite{sargsyanPhysLetA2021}. As the magnetic field is increased above 1\;kG, the intensity of the MI transition decreases. According to Fig.\,\ref{fig:Fig1}, the ratio of the amplitude of the MI transition \textcircled{7} at 1.30\;kG to that at 2.90\;kG should be about 1.8. However, on the spectra in Fig.\;\ref{fig:Fig4}, the observed ratio is about 2.7. This discrepancy is due to the increased transition broadening because of the magnetic field gradient from the PMs. The calculated spectra show a very good agreement with the experimental ones. To match, the broadening $\Gamma/2\pi$ in Eq.\;\ref{eq:profile} had to be increased from 150 to 210\;MHz. Note that the SD transmission spectrum is about 2.5 times narrower than the transmission spectrum given by Eq.\;\ref{eq:spectrumT}.



\section{Conclusion}

In conclusion, we have studied the evolution of $F_g \rightarrow F_g +2$ MI transitions of Rb and Cs D$_2$ lines, in the presence of a magnetic field upon linear $\pi$ excitation. In the case of $\pi$ excitation, the applied magnetic field is collinear with the laser polarization. In this configuration, the vapor is much more sensitive to the gradient produced by permanent magnets, which adversely affects the transition linewidth. This was partly overcome using weak-probe second derivative spectroscopy \cite{dutier2003collapse, sargsyanOL2019}, allowing individual atomic transitions to be resolved. For magnetic fields $B > 0.5 B_0$, we predict and observe that only one MI transition (out of possible 3 for $^{87}$Rb, 5 for $^{85}$Rb and 7 for $^{133}$Cs) remains in the spectrum. This is in contrast with the evolution of MI transitions in the case of circular $\sigma^+$ polarized excitation \cite{klinger2017magnetic,sargsyanPhysLetA2021}. 
Our calculations also show that such a feature exists for the MI transitions of all alkali atoms D$_2$ lines such as Na, K, Li, etc. Furthermore, a similar feature is expected for \(n^2\text{S}_{1/2}\rightarrow(n+1)^2\text{P}_{3/2}\) alkali lines, where $n= 3, 4, 5, 6$ is the principal quantum number for Na, K, Rb, Cs respectively, which amounts to about 100 MI transitions \cite{sargsyan2022dominant}. With the rapid development of micro-machined vapor-cell-based sensors, these results are important for magnetometers operating above Earth field, wide-range laser frequency stabilization systems, and atomic Faraday filters.





\section*{Funding}
 The work was supported by the Science Committee of RA, in the frame of the research project No 1-6/23-I/IPR.

\section*{Acknowledgments}
We thank Ara Tonoyan for fruitfull discussions.

\section*{Disclosures}
The authors declare no conflicts of interest.

\section*{Data availability statement}
The data that support the findings of this study are available from the corresponding author upon reasonable request.

\bibliography{Bib-main}

\end{document}